\documentstyle[aps,prd,preprint,epsf,floats]{revtex}
\tighten

\newcommand{\pp}{\psi(2S)}
\newcommand{\rt}{\rightarrow}
\newcommand{\ppj}{\psi(2S) \rightarrow
 \pi^+ \pi^- J/\psi}
\newcommand{\ppll}{\psi(2S) \rightarrow
 \pi^+ \pi^- J/\psi$, $J/\psi \rightarrow l^+ l^-}
\newcommand{\ppmm}{\psi(2S) \rightarrow
 \pi^+ \pi^- J/\psi$, $J/\psi \rightarrow \mu^+ \mu^-}
\newcommand{\ppee}{\psi(2S) \rightarrow
 \pi^+ \pi^- J/\psi$, $J/\psi \rightarrow e^+ e^-}

\newcommand{\etal}{\it et al.\rm}

\title{\boldmath $\psi(2S)\rightarrow \pi^{+}\pi^{-}J/\psi$ Decay
Distributions \cite{support}} 

\author{
J.~Z.~Bai,$^1$   Y.~Ban,$^5$      J.~G.~Bian,$^1$
I.~Blum,$^{12}$ 
G.~P.~Chen,$^1$  H.~F.~Chen,$^{11}$  
J.~Chen,$^3$ 
J.~C.~Chen,$^1$  Y.~Chen,$^1$ Y.~B.~Chen,$^1$
% ...
\cite{atNU0}
Y.~Y.~Wang,$^1$  
M.~Weaver,$^2$
C.~L.~Wei,$^1$   
N.~Wu,$^1$       Y.~G.~Wu,$^1$
D.~M.~Xi,$^1$    X.~M.~Xia,$^1$   P.~P.~Xie,$^1$    Y.~Xie,$^1$
Y.~H.~Xie,$^1$   G.~F.~Xu,$^1$    S.~T.~Xue,$^1$
J.~Yan,$^1$      W.~G.~Yan,$^1$   C.~M.~Yang,$^1$   C.~Y.~Yang,$^1$
H.~X.~Yang,$^1$  J.~Yang,$^1$     
W.~Yang,$^3$
X.~F.~Yang,$^1$  M.~H.~Ye,$^1$    S.~W.~Ye,$^{11}$
Y.~X.~Ye,$^{11}$ C.~S.~Yu,$^1$    C.~X.~Yu,$^1$     G.~W.~Yu,$^1$
Y.~H.~Yu,$^4$    Z.~Q.~Yu,$^1$    C.~Z.~Yuan,$^1$   Y.~Yuan,$^1$
B.~Y.~Zhang,$^1$ C.~Zhang,$^1$    C.~C.~Zhang,$^1$ D.~H.~Zhang,$^1$  
Dehong~Zhang,$^1$
H.~L.~Zhang,$^1$ J.~Zhang,$^1$    J.~W.~Zhang,$^1$  L.~Zhang,$^1$
L.~S.~Zhang,$^1$ P.~Zhang,$^1$
Q.~J.~Zhang,$^1$ S.~Q.~Zhang,$^1$ X.~Y.~Zhang,$^6$  Y.~Y.~Zhang,$^1$
D.~X.~Zhao,$^1$  H.~W.~Zhao,$^1$  Jiawei~Zhao,$^{11}$ J.~W.~Zhao,$^1$
M.~Zhao,$^1$     W.~R.~Zhao,$^1$  Z.~G.~Zhao,$^1$   J.~P.~Zheng,$^1$
L.~S.~Zheng,$^1$ Z.~P.~Zheng,$^1$ B.~Q.~Zhou,$^1$   G.~P.~Zhou,$^1$
H.~S.~Zhou,$^1$  L.~Zhou,$^1$     K.~J.~Zhu,$^1$    Q.~M.~Zhu,$^1$
Y.~C.~Zhu,$^1$   Y.~S.~Zhu,$^1$   B.~A.~Zhuang$^1$
\\ (BES Collaboration)}

\address{
$^1$Institute of High Energy Physics, Beijing 100039, People's Republic of
 China\\
$^2$California Institute of Technology, Pasadena, California 91125\\
$^3$Colorado State University, Fort Collins, Colorado 80523\\
$^4$Hangzhou University, Hangzhou 310028, People's Republic of China\\
$^5$Peking University, Beijing 100871, People's Republic of China\\
$^6$Shandong University, Jinan 250100, People's Republic of China\\
$^7$Shanghai Jiaotong University, Shanghai 200030, People's Republic of China\\
$^8$Stanford Linear Accelerator Center, Stanford, California 94309\\
$^9$University of Hawaii, Honolulu, Hawaii 96822\\
$^{10}$University of California at Irvine, Irvine, California 92717\\
$^{11}$University of Science and Technology of China, Hefei 230026,
People's Republic of China\\
$^{12}$University of Texas at Dallas, Richardson, Texas 75083-0688}

\date{\today}
\begin{document}
\maketitle

\begin{abstract}
Using a sample of 3.8 M $\pp$ events accumulated with the BES
detector, the process $\ppj$ is studied.  The angular distributions
are compared with the general decay amplitude analysis of Cahn.  We
find that the dipion system requires some D-wave, as well as S-wave.
On the other hand, the $J/\psi$ - $(\pi^+ \pi^-)$ relative angular momentum is
consistent with being pure S-wave.  The decay distributions have been fit
to heavy quarkonium models, including the Novikov-Shifman model.
This model, which is written in terms of the parameter $\kappa$,
predicts that D-wave should be present.  We determine $\kappa = 0.183
\pm 0.002 \pm 0.003$ based on the joint $m_{\pi \pi}$ - $\cos
\theta_{\pi}^*$ distribution.  The fraction of D-wave as a function of
$m_{\pi \pi}$ is found to decrease with increasing $m_{\pi \pi}$, in
agreement with the model.  We have
also fit the Mannel-Yan model, which is another model that allows D-wave.

\end{abstract}




\section{Introduction}

Transitions between bound $c \bar{c}$ states as well as between $b
\bar{b}$ states provide an excellent laboratory for studying heavy
quark-antiquark dynamics at short distances.  Here we study 
the process $\ppj$, which  is the largest decay mode of the
$\psi(2S)$ \cite{pdg}.
The dynamics of this process 
can be investigated using very clean exclusive $\ppll$ events, where
$l$ signifies either $e$ or $\mu$.

Early investigation of this decay
by Mark I \cite{abrams}
found that the $\pi^+ \pi^-$ mass distribution was
strongly peaked towards higher mass values, in contrast to what was
expected from phase space.  Further, angular distributions
strongly favored  S-wave production of $\pi \pi J/\psi$, as well as an
S-wave decay of the dipion system.

The challenge of describing the mass spectrum attracted considerable
theoretical interest.  Brown and Cahn \cite{brown} and Voloshin
\cite{voloshin2} used chiral symmetry arguments and partially conserved
axial currents (PCAC) to derive a
matrix element. 
Assuming chiral symmetry breaking to be small, Brown and Cahn showed the
decay amplitude for this process involves three parameters, which are
the coefficients of three different momentum dependent terms.  If two
of the parameters vanish, then the remaining term would give a peak
at high invariant mass along
with the isotropic S-wave behavior.  In a more general analysis, Cahn
\cite{cahn} calculated the angular distributions in terms of partial
wave amplitudes diagonal in orbital and spin angular momentum.

\begin{figure}[!htb]
\centerline{\epsfysize 1.8 truein
\epsfbox{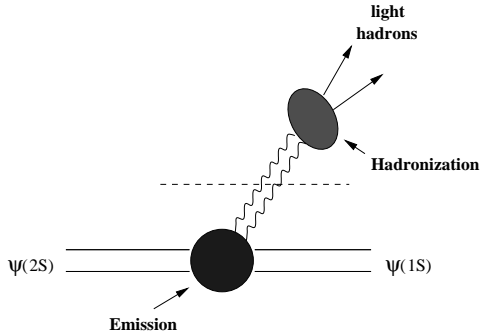}}
\caption{\label{fig:zijin}
Diagram of $\ppj$ decay process, showing it as a two step process
with the
emission of two gluons followed by hadronization to pion pairs.
}
\end{figure}

These transitions are thought to occur in a two step process by the
emission of two gluons followed by hadronization to pion pairs, as
indicated in Fig.~\ref{fig:zijin}.
Because of the small mass difference involved, the gluons are soft and
can not be handled by perturbative QCD.  However, Gottfried
\cite{gottfried} suggested that the gluon emission can be described by a
multipole expansion with the gluon fields being
expanded in a multipole series similar to electromagnetic transitions.
Including the leading chromo-electric $E1$$E1$ transition, T. M. Yan \cite{yan1}
determined that one of the terms that Brown and Cahn took to be zero should
have a small but nonzero value.
Voloshin and Zakharov \cite{Voloshin} and later, in a revised
analysis, Novikov and Shifman \cite{vzns} worked out the second step, the
pion hadronization
matrix element using current
algebra, PCAC, and gauge invariance.  They were able to derive an
amplitude for this process from ``first principles.''
Interestingly, Ref. \cite{vzns} predicts that while the decay should be
predominantly S-wave, a small amount of D-wave should be present in
the dipion system.

All models predict the spectrum to peak at high mass as it does in
$\ppj$ and $\Upsilon(2S) \rt \pi \pi \Upsilon(1S)$.  However,
$\Upsilon(3S) \rt \pi^+ \pi^- \Upsilon(1S)$ \cite{upsilon3s}
has a peak
at low mass, as well as a peak at high mass, that disagrees with
these predictions.  See Ref. \cite{upsilon3s} for a list of theory papers
that attempt to deal with this problem.

In this paper, we will study the decay distributions of the
$\ppj$ process and use them to test models.  The events come from a data
sample of $3.8 \times 10^6$ $\psi(2S)$ decays taken with the BES detector.

\section{The BES detector}

The Beijing Spectrometer, BES,
is a conventional cylindrical magnetic detector that is coaxial
with the BEPC colliding $e^+e^-$ beams.  It is 
described in detail in Ref.~\cite{bes}. A four-layer central drift
chamber (CDC) surrounding the beampipe provides trigger
information. Outside the CDC, the forty-layer main drift chamber (MDC)
provides tracking and energy-loss ($dE/dx$) information on
charged tracks over $85\%$ of the total solid angle.
The momentum resolution is $\sigma _p/p = 1.7 \% \sqrt{1+p^2}$ ($p$
in GeV/c), and the $dE/dx$ resolution for hadron tracks for this
data sample is $\sim 9\%$. 
An array of 48 scintillation counters surrounding the MDC provides 
measurements of the time-of-flight (TOF) of charged tracks with a resolution of
$\sim 450$ ps for hadrons. Outside the TOF system, a 12
radiation length lead-gas barrel shower counter (BSC),
operating in self-quenching streamer mode, measures the energies 
of electrons and photons over  80\% of the total solid
angle. The energy resolution is $\sigma_E/E= 22 \%/\sqrt{E}$ ($E$
in GeV).
Surrounding the BSC is a solenoidal magnet that
provides a 0.4 Tesla magnetic field in the central tracking
region of the detector. Three double layers of proportional chambers
instrument the magnet flux return (MUID) and are used to identify
muons of momentum greater than 0.5 GeV/c.

\section{Event Selection}

In order to study the process $\ppj$, we use the very clean $\ppll$
sample. The initial event selection is the same as in Ref.~\cite{bll}.
We require four tracks total with the sum of the charge equal zero.

\subsection{Pion Selection}

We require a pair of oppositely charged candidate
pion tracks with good helix fits that satisfy:
\begin{enumerate}
\item $|\cos \theta_{\pi}| < 0.75$.
Here $\theta_{\pi}$ is the polar angle of the $\pi$ in the laboratory system.
\item $p_{\pi} < 0.5$ GeV/c,
where $p_{\pi}$ is the pion momentum. 
\item $pxy_{\pi} > 0.1$ GeV/c,
where $pxy_{\pi}$ is the momentum of the pion transverse to the beam
direction.  This removes tracks that circle in the Main Drift Chamber.
\item $\cos \theta_{\pi \pi} < 0.9, $
where $\theta_{\pi \pi}$ is the laboratory angle between the $\pi^+$ and $\pi^-$.
This cut is used to eliminate contamination from misidentified
$e^+e^-$ pairs from $\gamma$ conversions.
\item $3.0 < m_{recoil} < 3.2$ GeV/c$^2$, where $m_{recoil}$ is the mass
recoiling against the dipion system.

\item $|\chi^{dE/dx}_{\pi}| < 3.0$. $\chi^{dE/dx}_{\pi} = \frac{(dE/dx)_{meas}
 - (dE/dx)_{exp}}{\sigma}$
where $(dE/dx)_{meas}$ and $(dE/dx)_{exp}$ are the measured
and expected $dE/dx$ energy
losses 
for pions, respectively, and $\sigma$ is the experimental $dE/dx$ resolution.
\end{enumerate} 

\subsection{Lepton Selection}
The lepton tracks must satisfy:
\begin{enumerate}

\item $0.5 < p_l < 2.5$ GeV/c.
Here $p_l$ is the three-momentum of
the candidate lepton track.

\item
$|\cos\theta_{e}|<0.75$, $|\cos\theta_{\mu}|<0.60$.
Here $\theta_{e}$ and $\theta_{\mu}$ are the laboratory polar angles of the
electron and muon, respectively.
This cut ensures that 
electrons are contained in the $BSC$ and muons in
the MUID system.

 
\item $\cos \theta_{l^+ l^-}^{cm} < -0.975$. This is the cosine of the angle
between the two leptons in the $J/\psi$ CM, where
the leptons are nearly back-to-back.

\item $p_{l^+}$ or $p_{l^-} > 1.3$ GeV/c or $p_{l^+} + p_{l^-} > 2.4$ GeV/c.
This cut
selects events consistent with $J/\psi$ decay, while rejecting
background. 

\item 
For $e^+e^-$ candidate pairs: $SCE_+$ and $SCE_- >
0.6$ GeV/c, 
where $SCE$ is the energy deposited in the BSC, or, if one of the tracks goes
through a BSC rib or has $P_l < 0.8 $ GeV/c, the $dE/dx$ information of
both tracks in the MDC must
be consistent with that expected for electrons. The rib region of the
BSC is not used because the Monte Carlo does not model the energy deposition
well in this region. 

\item 
For $\mu^{+}\mu^{-}$ pair candidates at least one track must have
$N^{hit} > 1$,
where $N^{hit}$ is the number of MUID layers with matched hits 
and ranges from 0 to 3. If only one track is identified in this fashion, then
the invariant mass of the $\mu \mu$ pair must also be within 250 MeV/c$^2$ of
the $J/\psi$ mass. 

\end{enumerate}

\subsection{Additional Criteria}
\begin{figure}[!htb]
\centerline{\epsfysize 2.75 truein
\epsfbox{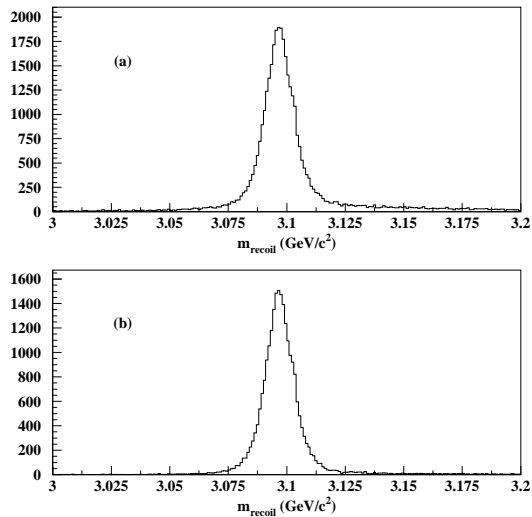}}
\caption{\label{fig:mrecoil}
{\bf (a)} Number of events versus
$m_{recoil}$, the mass recoiling against the two $\pi$'s, for
$\psi(2S) \rightarrow \pi^+ \pi^- J/\psi, J/\psi \rightarrow l^+ l^-$
events where only the initial selection criteria (see text) are used.
{\bf (b)} Number of events versus $m_{recoil}$ 
where all criteria are used except for the final $m_{recoil}$ cut.
}
\end{figure}

Fig.~\ref{fig:mrecoil}a shows the $m_{recoil}$ distribution using the
cuts defined above.  The shoulder above the $J/\psi$ peak is caused by
low energy pions that undergo $\pi \rt \mu \nu$ decay.  We impose
additional selection criteria in order to reduce the amount of
mismeasured events from these and from other events where the $J/\psi$
undergoes final state radiation or where electrons radiate much of
their energy.  These cuts are necessary for comparisons with
theoretical models.

\begin{enumerate}
\item The $\pi$'s must be consistent with coming from the interaction
point.
\item $3.07 < m_{recoil} < 3.12$ GeV/c$^2$.
\item $| m_{l^+ l^-} - m_{J/\psi}| < 0.25$ GeV/c$^2$, where $m_{l^+
    l^-}$ is the invariant mass of the two leptons.
\end{enumerate}
Fig.~\ref{fig:mrecoil}b shows the $m_{recoil}$ distribution using all
cuts except the
additional $m_{recoil}$ cut (additional cut number 2).
A total of 22.8 K events remains after all cuts, and
the background
remaining is estimated to be less than 0.3 \%.

\section{Monte Carlo}
The process is considered to take place via sequential 2-body decays:
$\psi(2S) \rightarrow X + J/\psi$,
$X \rightarrow \pi^+ \pi^-$, and $J/\psi \rightarrow l^+ l^-$.
The Monte Carlo program assumes:
\begin{enumerate}
\item The mass of the dipion system is empirically given by:
\[ \frac{d \sigma}{dm_{\pi \pi}} \propto ({\rm phase } \, {\rm space }) \times
(m_{\pi \pi}^2 - 4 m_{\pi}^2)^2 \]
\item The orbital angular momentum between the dipion system and
the $J/\psi$ and between the $\pi$'s in the $\pi^+\pi^-$ system is 0.
\item The X and the $J/\psi$ are uniformly distributed in $\cos \theta$ in the
incoming $e^+ e^-$ rest frame, which is the same as the laboratory
frame.
\item The $\pi$'s are uniformly distributed in $\cos \theta_{\pi}^*$,
where $\theta_{\pi}^*$ is the angle between the $J/\psi$ direction
and the $\pi^+$  in the X rest
frame.
\item Leptons have a $1 + \cos^2 \theta_l^*$ distribution,
where $\theta_l^*$ is the angle between the beam direction and the
positive lepton in the $J/\psi$ rest frame.

\item The $J/\psi$ decay has an order $\alpha^3$ final state radiative
correction in the rest frame of the $J/\psi$.
\end{enumerate}

A total of 570,000 Monte Carlo events are
generated each for the $\ppee$ and $\ppmm$ samples.


In order to compare with theoretical models, the experimental
distributions must be corrected for detection efficiency.  To
determine this correction, Monte Carlo data is run through the same analysis
program as the data.  A bin-by-bin efficiency correction is then determined for
each distribution of interest using the generated and detected Monte
Carlo data.  This efficiency is then used to correct each
bin of the data distributions \cite{binbybin}.

\begin{figure}[!htb]
\centerline{\epsfysize 4.0 truein
\epsfbox{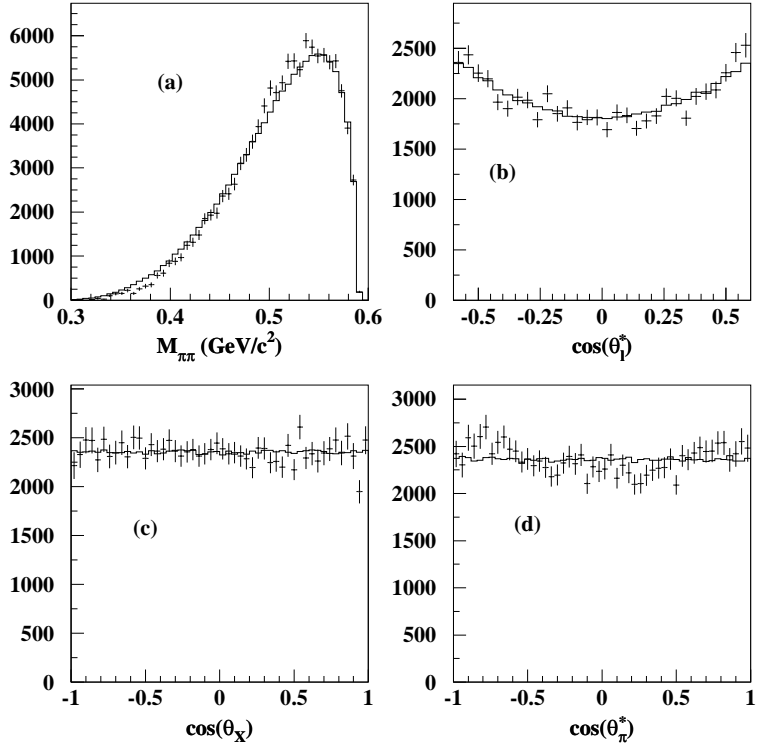}}
\caption{\label{fig:mccompare}
Various distributions (corrected for detection efficiency)
for $\ppll$ decays. {\bf (a)} $m_{\pi^+ \pi^-}$
distribution.  The distribution is in reasonable
agreement with the assumed
empirical distribution.  {\bf (b)} $\cos \theta_l^*$ distribution.  The
assumed distribution is a $1 + \cos^2 \theta_l^*$ distribution.
This angle is the angle between the beam direction and the $l^+$ in the
rest frame of the $J/\psi$.
{\bf (c)} $\cos
\theta_{X}$ distribution.  This is the cosine of the
angle of the dipion system
with respect to the $e^+ e^-$ direction in the incoming $e^+ e^-$ CM
system.
The
distribution for Monte Carlo data is flat because of the S-wave assumption
for the relative
angular momentum of the dipion system and the $J/\psi$.  {\bf (d)} $\cos
\theta_{\pi^+}^*$ distribution.  This is the cosine of the angle
of the $\pi^+$ with
respect to the $J/\psi$ direction in the dipion rest frame.  The
Monte Carlo distribution is flat because of the assumption that the
relative angular momentum of the $\pi$'s is S-wave. The data agree
well with the Monte Carlo except in (d).
}
\end{figure}

A comparison of some distributions with the Monte Carlo distributions
is shown in Fig.~\ref{fig:mccompare}.
Fig.~\ref{fig:mccompare}a indicates that the $m_{\pi \pi}$
distribution agrees qualitatively with the assumed empirical
distribution \cite{mass_res}.  Fig.~\ref{fig:mccompare}b indicates agreement with the
assumed $1 + \cos^2 \theta_l^*$ distribution for leptons in $\ppll$
events.  The flat distribution in Fig.~\ref{fig:mccompare}c is related to
the assumption that the relative angular momentum between the dipion
system and the $J/\psi$ is zero. 
However, in Fig.~\ref{fig:mccompare}d, which is the
$\cos
\theta_{\pi^+}^*$ distribution, we find a disagreement with the Monte
Carlo data, indicating that the 
relative angular momentum of the two $\pi$'s is inconsistent with being pure
S-wave.

\begin{figure}[!htb]
\centerline{\epsfysize 4.0 truein
\epsfbox{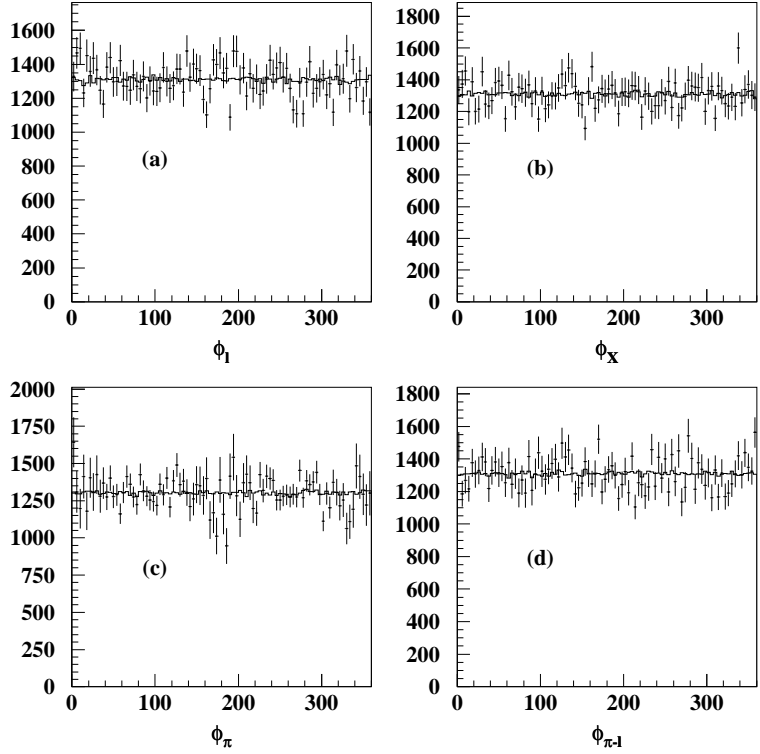}}
\caption{\label{fig:phi}
Azimuthal angle distributions (corrected for detection efficiency)
for $\ppll$ decays. {\bf (a)}
The $\phi$ angle distribution for the $l^+$ in the lab.
{\bf (b)} The $\phi$ angle distribution for $X$ in the lab.
{\bf (c)} The $\phi$ angle distribution for the $\pi^+$ in the dipion
rest frame.
{\bf (d)} The distribution of the angle
between the normals to the $\mu \mu$ plane and the $\pi \pi$ plane.
}
\end{figure}

Fig.~\ref{fig:phi} shows the $\phi$ angle distributions for the
$l^+$ in the lab; the $J/\psi$ in the lab; the $\pi^+$ in the
rest frame of the dipion system, $\phi_{\pi^+}$; and the angle between
the normals to the $\mu \mu$ plane and the $\pi \pi$ plane.
\[ \phi_{\pi^+} = \arctan \left[\frac{(\widehat{(\widehat{\hat{X} \times
\hat{z})} \times \hat{X})} \cdot \widehat{p_{\pi^+}}}{\widehat{(\hat{X}
\times \hat{z})} \cdot \widehat{p_{\pi^+}}}\right] \]
All distributions are uniform in angle,
consistent with the Monte Carlo distributions.

Since Fig.~\ref{fig:mccompare}d indicates an inadequacy with the Monte
Carlo, it is necessary to correct our bin-by-bin efficiency
determination in our following studies.  We use the
Novikov-Shifman model (discussed below), which gives a reasonable
approximation to the data,  to determine a weighting for
Monte Carlo events so that the proper efficiency is determined as a
function of $\cos \theta_{\pi}^*$ and $m_{\pi \pi}$.

\section{\boldmath Angular Distributions and
Partial Wave Analysis}
In this section, we fit our angular distributions using
the general decay amplitude analysis of Cahn \cite{cahn}.
The $\psi(2S)$ and $J/\psi$ have $J^P = 1^-$ and $I^{GC} = 0^{--}$,
while the dipion system has $I^{GC} = 0^{++}$.
At an $e^+ e^-$ machine, the
$\psi(2S)$ is produced with  
polarization transverse to the beam.
The decay of $\pp$
can be described by the quantum numbers:
\begin{description}
\item [\boldmath $\vec{l}$] $~-~\pi \pi$ angular momentum
\item [\boldmath $\vec{L}$] $~-~J/\psi$ X angular momentum
\item [\boldmath $\vec{s}$]  $~-~$ spin of the $J/\psi$
\item [\boldmath $\vec{s}{~'}$]  $~-~$ spin of the $\pp$
\end{description}
Defining $\vec{S} = \vec{s} + \vec{l}$, called the channel spin,
then $\vec{s}^{~'} = \vec{S} + \vec{L} = \vec{s} + \vec{l} + \vec{L}$.
An eigenstate of $J^2 = s^{'2}$, $L^2$, $S^2$, and $J_z$ may be constructed.
Parity conservation and charge
conjugation invariance require both $L$ and $l$ to be even.

The decay can be described in terms of partial wave
amplitudes, $M_{l,L,S}$, and the partial waves can be truncated after a
few terms.
Considering only $M_{001}$, $M_{201}$, and
$M_{021}$ \cite{cahnerror}:
\begin{equation}
\frac{d \Gamma}{d \Omega_{J/\psi}} \propto [ |M_{001}|^2 + |M_{201}|^2
+ \frac{1}{4}|M_{021}|^2(5 - 3 \cos^2 \theta_{J/\psi}^*)
+ \frac{1}{\sqrt 2}\Re \{M_{021}M_{001}^*\}(3\cos^2
\theta_{J/\psi}^*-1)] \label{eqn:cahn1'}
\end{equation}
\begin{equation}
\frac{d \Gamma}{d \Omega_{\pi}} \propto [ |M_{001}|^2
 + \frac{1}{4}|M_{201}|^2(5 - 3\cos^2 \theta_{\pi}^*)
 + |M_{021}|^2
+ \frac{1}{\sqrt 2}\Re \{M_{201}M_{001}^*\}(3\cos^2 \theta_{\pi}^*-1)]
\label{eqn:cahn2'}
\end{equation}
\begin{equation}
\frac{d \Gamma}{d \Omega_{\mu}} \propto [ |M_{001}|^2(1 
+ \cos^2 \theta_{\mu}^*) +\frac{1}{10}(|M_{201}|^2 + |M_{021}|^2)(13
+ \cos^2 \theta_{\mu}^*)] \label{eqn:cahn3}
\end{equation}
The $d \Omega$'s are measured in their respective rest frames.
It is
understood that the $M_{l,L,S}$ are functions of $m_{\pi \pi}$.  
The combined $\theta_{\pi} - \theta_{J/\psi}$ distribution is given by:
\clearpage
\begin{eqnarray}
\frac{d \Gamma}{d \Omega_{\pi} d \Omega_{J/\psi}} & \propto & |M_{001}|^2
+ |M_{201}|^2\left(\frac{5}{4} - \frac{3}{4}\cos^2 \theta_{\pi}^*\right)
+ |M_{021}|^2
\left(\frac{5}{4} - \frac{3}{4} \cos^2 \theta_{J/\psi}^*\right) \nonumber \\
  & & +2 \Re \{M_{201}M_{001}^*\}\left[\frac{1}{\sqrt 2}\left(\frac{3}{2}\cos^2 \theta_{\pi}^*
- \frac{1}{2}\right)\right] \nonumber \\
  & & + 2 \Re \{M_{021} M_{001}^*\} \left[\frac{1}{\sqrt 2}\left(\frac{3}{2}
\cos^2 \theta_{J/\psi}^* - \frac{1}{2}\right)\right]    \nonumber     \\
  & & + 2 \Re \{M_{201} M_{021}^*\}\left[\frac{9}{8}\sin^2 \theta_{\pi}^* \sin^2
\theta_{J/\psi}^* \cos 2(\phi_{\pi}^* - \phi_{J/\psi}^*) \right. \nonumber \\
 & & +\frac{9}{16}
\sin 2 \theta_{\pi}^* \sin 2 \theta_{J/\psi}^* \cos(\phi_{\pi}^* - \phi_{J/\psi}^*) \nonumber \\
  & & + \left. \frac{1}{2}\left(\frac{3}{2} \cos^2 \theta_{\pi}^* - \frac{1}{2}\right)
\left(\frac{3}{2}\cos^2 \theta_{J/\psi}^* - \frac{1}{2}\right) \right] \label{eqn:cahn2d}
\end{eqnarray}

\subsection{Fits to 1D Angular Distributions}
\begin{figure}[!htb]
\centerline{\epsfysize 4.0 truein
\epsfbox{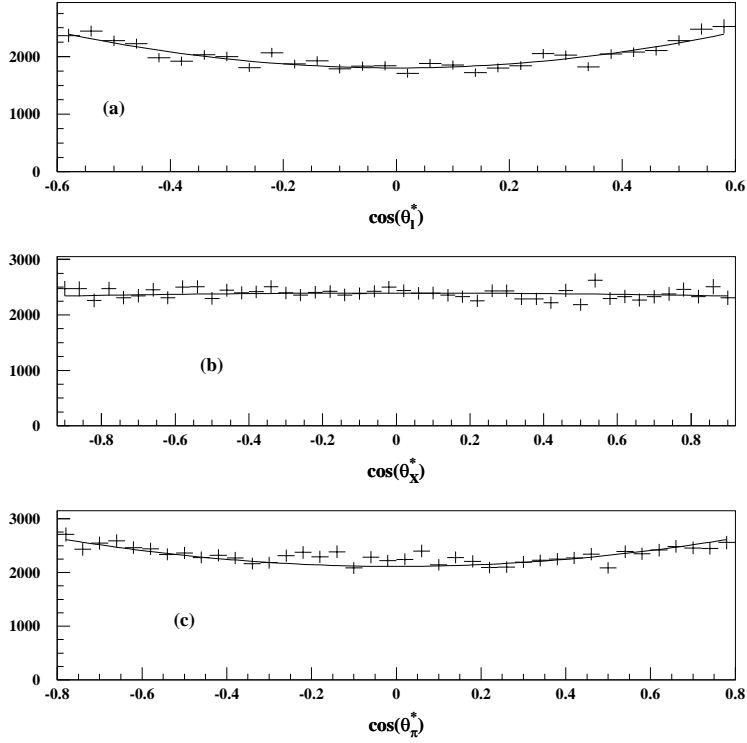}}
\caption{\label{fig:simul1da}
Simultaneous fits ($\chi^2$) to three 1-D histograms of
{\bf (a)} $\cos \theta_{\mu}^*$, {\bf (b)} $\cos \theta_X^*$, and
{\bf (c)} $\cos \theta_{\pi}^*$
using Eqns.~\ref{eqn:cahn3}, \ref{eqn:cahn1'}, and \ref{eqn:cahn2'},
respectively.
The phase shifts used are $\delta^0_0 =  45^{\circ}$ 
and $\delta^0_2 = 0^{\circ}$.  
}
\end{figure}

There are three complex numbers to be obtained.
According to Cahn, if the $\psi(2S)$ and $J/\psi$ are regarded as inert, then
the usual final state argument gives $M_{l,L,S} = e^{i\delta^0_l(m_{\pi \pi})}
|M_{l,L,S}|$, where $\delta^0_l(m_{\pi \pi})$ is the isoscalar phase shift
for quantum number $l$. 
The phase angles are functions of
$m_{\pi \pi}$.
If we interpolate the S wave, isoscalar phase
shift data
found in Ref.~\cite{Belanger}, we find
$\delta^0_0 \approx 45^{\circ}$.  Also $\delta^0_2$ is supposed to be $\approx 0$.
Using these values as input, we obtain the combined fit to
Eqns.~\ref{eqn:cahn1'}-\ref{eqn:cahn3}, shown in
Fig.~\ref{fig:simul1da} \cite{fit_range}, and the results given in
Table~\ref{table:simul1da}. Also
given in Table~\ref{table:simul1da} are the ratios $|M_{201}|/|M_{001}|$
and $|M_{021}|/|M_{001}|$. 
The fit yields a nonzero result for
$|M_{201}|$, indicating that the dipion system contains some D-wave.
The amplitude $|M_{021}|$ is very small, indicating that the $J/\psi X$
angular momentum is consistent with zero.   

Cahn points out that one of the advantages of the
process $\ppj$ is that
it may allow us to obtain $\delta^0_0$, which is not well measured in this mass
range.
However we are unable to obtain a good fit allowing $\delta^0_0$ as an
additional parameter.

\begin{table}[!h]
\caption{Results of simultaneous $\chi^2$ fits to the three 1-D distributions of
$\cos \theta_{\mu}^*$, $\cos \theta_X^*$, and $\cos \theta_{\pi}^*$
shown in Fig.~\ref{fig:simul1da}.
The phase shifts used are $\delta^0_0 =  45^{\circ}$ and
$\delta^0_2 = 0^{\circ}$. The amplitude normalizations are arbitrary.
Two other fit
parameters (not shown) are the normalizations of the second
and third distributions relative to the first. \label{table:simul1da}
}
\

\begin{center}
\begin{tabular}{ll}
$|M_{001}|$   & $41.6 \pm 0.4 \pm 0.9$      \\
$|M_{201}|$   & $7.5 \pm 1.4 \pm 1.9$     \\
$|M_{021}|$   & $-0.56 \pm 0.60 \pm 0.64$     \\
$|M_{201}|/|M_{001}|$ & $0.18 \pm 0.03 \pm 0.05$   \\
$|M_{021}|/|M_{001}|$ & $-0.013 \pm 0.014 \pm 0.015$  \\
$\chi^2/DOF$ & 89/111                        \\
\end{tabular}
\end{center}
\end{table}

\subsection{Fits to the 2D Distribution}
By integrating Eqn.~\ref{eqn:cahn2d} over the $\phi$ angles, we obtain an expression that depends
only on $\cos \theta_{\pi}^*$ and $\cos \theta_{J/\psi}^*$:
\newline
\begin{minipage}[t]{6.5 in}
\begin{eqnarray}
\frac{d \Gamma}{d \cos \theta_{\pi}^*  d \cos \theta_{J/\psi}^*} & \propto & |M_{001}|^2
+ |M_{201}|^2\left(\frac{5}{4} - \frac{3}{4}\cos^2 \theta_{\pi}^*\right)
+ |M_{021}|^2
\left(\frac{5}{4} - \frac{3}{4} \cos^2 \theta_{J/\psi}^*\right) \nonumber \\
  & & +2 \Re \{M_{201}M_{001}^*\}\left[\frac{1}{\sqrt 2}\left(\frac{3}{2}\cos^2 \theta_{\pi}^*
- \frac{1}{2}\right)\right] \nonumber \\
  & & + 2 \Re \{M_{021} M_{001}^*\} \left[\frac{1}{\sqrt 2}\left(\frac{3}{2}
\cos^2 \theta_{J/\psi}^* - \frac{1}{2}\right)\right]    \nonumber     \\
  & & + \Re \{M_{201} M_{021}^*\}\left[
  \left(\frac{3}{2} \cos^2 \theta_{\pi}^* - \frac{1}{2}\right)
\left(\frac{3}{2}\cos^2 \theta_{J/\psi}^* - \frac{1}{2}\right)\right] \label{eqn:cahn2dnophi}
\end{eqnarray}
\end{minipage}
The 2D distribution of $\cos \theta_{\pi}^*$ versus
$\cos \theta_{J/\psi}^*$ is fit using this equation.  We assume
$\delta^0_0 = 45^{\circ}$ and $\delta^0_2 = 0^{\circ}$, as was done previously.
Using these values, we obtain the fit values shown in Table~\ref{tab:cahn3}.
If we try to obtain $\delta^0_0$,
we are unable to get a good fit.

Fits for different $m_{\pi \pi}$
intervals are made assuming $\delta^0_2 = 0^{\circ}$ and using values
of $\delta^0_0$ that depend on the $m_{\pi \pi}$ interval\cite{delta00}.
The results
are shown in Table~\ref{tab:interval}, along with the values of $\delta^0_0$
used.  The ratios $|M_{201}|/|M_{001}|$
and $|M_{021}|/|M_{001}|$ do not show large variations between the three
intervals,  and
$|M_{201}|/|M_{001}|$ is inconsistent with zero for all intervals.

\begin{table}[!h]
\caption{Result of the 2-D Likelihood fit to $\cos \theta_{\pi}^*$ versus $\cos \theta_X^*$
using Eqn.~\ref{eqn:cahn2dnophi}.
The phase shifts used are $\delta^0_0 =  45^{\circ}$ and
$\delta^0_2 = 0^{\circ}$.  The amplitude normalizations are arbitrary. 
}
\
\begin{center}
\label{tab:cahn3}
\begin{tabular}{ll}
$|M_{001}|$         & $13.6 \pm 0.05 \pm 0.25$  \\
$|M_{201}|$         & $2.3  \pm 0.3 \pm 0.5$   \\
$|M_{021}|$         & $0.05 \pm 0.16 \pm 0.19 $    \\
$|M_{201}|/|M_{001}|$ & $0.17 \pm 0.02 \pm 0.04$   \\
$|M_{021}|/|M_{001}|$ & $0.004 \pm 0.01 \pm 0.01$   \\
$\chi^2/DOF$ & 457/437                 \\
\end{tabular}
\end{center}
\end{table}

\begin{table}[!h]
\caption{Results of the 2-D Likelihood fits to $\cos \theta_{\pi}^*$
versus $\cos \theta_X^*$
using Eqn.~\ref{eqn:cahn2dnophi} for different $m_{\pi \pi}$ intervals.
The amplitude normalization is arbitrary. Here the value of $\delta^0_0$
used depends on the  $m_{\pi \pi}$ interval. $\delta^0_2 = 0$.
}
\
\begin{center}
\begin{tabular}{lccc}
~$m_{\pi \pi}$ Range (GeV/c$^2)$~&~~ 0.36 - 0.5~~&~0.5 - 0.54~~&~~0.54 - 0.6 \\
\label{tab:interval}
$\delta^0_0$ used as input   & $27^{\circ}$  & $42^{\circ}$ & $51^{\circ}$   \\ \hline
$|M_{001}|$   & $8.36 \pm 0.04 \pm 0.23$  & $7.43 \pm 0.04 \pm 0.14$ & $7.86 \pm 0.06 \pm 0.15$   \\
$|M_{201}|$   & $1.19 \pm 0.27 \pm 0.52$  & $0.89 \pm 0.29 \pm 0.28$ & $1.37 \pm 0.37 \pm 0.56$\\
$|M_{021}|$   & $0.53 \pm 0.19 \pm 0.43$  & $-0.27 \pm 0.15 \pm 0.18$ & $0.14 \pm 0.15 \pm 0.21$\\
$|M_{201}|/|M_{001}|$ & $0.14 \pm 0.03 \pm 0.06$ & $0.12 \pm 0.04 \pm 0.04$ & $0.17 \pm 0.05 \pm 0.07$\\
$|M_{021}|/|M_{001}|$ & $0.06 \pm 0.02 \pm 0.05 $ & $-0.04 \pm 0.02 \pm 0.02$& $0.02 \pm 0.02 \pm 0.03$\\
$\chi^2/DOF$ & 514/437                  & 608/437             &  545/437\\
Events         &  6186             & 7075      & 9362  \\ 
\end{tabular}
\end{center}
\end{table}

In comparing the results from
Tables~\ref{table:simul1da}-\ref{tab:interval}, we
see that $|M_{201}|/|M_{001}|$ varies between 0.12 and 0.18 and is at
least two sigma from zero.  On the other hand,
$|M_{021}|/|M_{001}|$ varies between -0.04 and 0.06 and is, in all cases,
consistent with zero.

\section{
Comparison with Heavy Quarkonium Models}

\subsection{Novikov-Shifman Model}
A model that predicts some D-wave is
the Novikov-Shifman
\cite{vzns} model, which is based on the color field multipole
expansion to describe the two gluon emission and uses chiral symmetry,
current algebra, PCAC, and gauge invariance to obtain the
matrix element.  In this model the transition is dominated by $E1E1$
gluon radiation, so the angular momentum of the $c \bar{c}$ system is
not expected to change during the decay and the polarization of the
$\pp$ should be the same as the $J/\psi$.  
\clearpage
\noindent The model gives the amplitude

\begin{eqnarray}
A & \propto & \left\{q^2 - \kappa(\Delta M)^2\left(1 + \frac{2 m^2_{\pi}}{q^2}\right) \right. \nonumber\\
& & + \frac{3}{2} \kappa \left.[(\Delta M)^2 - q^2]\left(1 - \frac{4 m^2_{\pi}}{q^2}\right)
\left(\cos^2 \theta_{\pi}^* - \frac{1}{3}\right)\right\}, \label{eq:a}
\end{eqnarray}
where q is the four momentum of the dipion system and $\Delta M = M_{\psi(2S)} -
M_{J/\psi}$.
The parameter $\kappa$ is given by
\begin{equation}
\kappa = (b/6 \pi) \alpha_s(\mu) \rho^G(\mu),\label{eqn:kappa}
\end{equation}
where $b = 9$ is the first expansion coefficient of the Gell-Mann-Low
function, $\rho^G$ is the gluon fraction of the $\pi$'s momentum, which
is about 0.4, and $\kappa$ is predicted to be $\approx 0.15$ to $0.2$
\cite{shifman}. From Eqn.~\ref{eqn:kappa}, it can be seen that $\kappa$
is expected to be different for $\psi(2S)$
decays and the decays of other charmonia, because of the running of $\alpha_s$.
The
first terms in the amplitude are the S-wave contribution, and the last
term is the D-wave contribution.
Note that parity and charge conjugation invariance
require that the spin be even.  If $\kappa$ is non-zero, it is
predicted that there should be some D-wave.  However, since $\kappa$ is
expected to be small, the process should be predominantly S-wave.

The differential cross section is obtained by squaring the amplitude and 
multiplying by phase space.
\begin{equation}
\frac{d \Gamma}{d m_{\pi \pi} d \cos \theta^*_{\pi}} \propto ({\rm PS}) \times A^2, \label{eqn:dif}
\end{equation}
where

\[ PS = \sqrt{\frac{(m_{\pi \pi}^2 -4 m_{\pi}^2)[M_{J/\psi}^4 + M_{\psi(2S)}^4 + m_{\pi \pi}^4
 -2(M_{J/\psi}^2m_{\pi \pi}^2 + M_{\psi(2S)}^2 m_{\pi \pi}^2 +M_{J/\psi}^2  M_{\psi(2S)}^2)]}{4 M_{\psi(2S)}^2}}  \]
By integrating over one variable at a time, it is possible to obtain
the
following 1D equations for the $m_{\pi \pi}$ invariant mass spectrum and
the $\cos \theta_{\pi}^*$ distribution:
\begin{eqnarray}
\frac{d \sigma}{d m_{\pi \pi}} & \propto & |\vec{q}|\sqrt{(q^2 - 4m^2_{\pi})}
\left\{\left[q^2 -
\kappa(\Delta M)^2\left(1 + \frac{2 m_{\pi}^2}{q^2}\right)\right]^2 \right. \nonumber  \\ \label{eqn:f1}
& & +\left. 0.2\kappa^2[(\Delta M)^2 - q^2]^2\left(1 -\frac{4m_{\pi}^2}{q^2}
\right)^2\right\} \\
\frac{d \sigma}{d \cos \theta_{\pi}^* } & \propto & \left\{1.322 - 4.8597\kappa + 5.1577 \kappa^2
+1.18296 \kappa\left(\cos^2 \theta^*_{\pi} - \frac{1}{3}\right) \right. \nonumber  \\
 & & - 2.65421 \left. \kappa^2\left(\cos^2 \theta^*_{\pi} - \frac{1}{3}\right)
+ 0.3738 \kappa^2\left(\cos^2 \theta^*_{\pi}-\frac{1}{3}\right)^2\right\} \label{eqn:g1}
\end{eqnarray}

\begin{figure}[!htb]
\centerline{\epsfysize 3.5 truein
\epsfbox{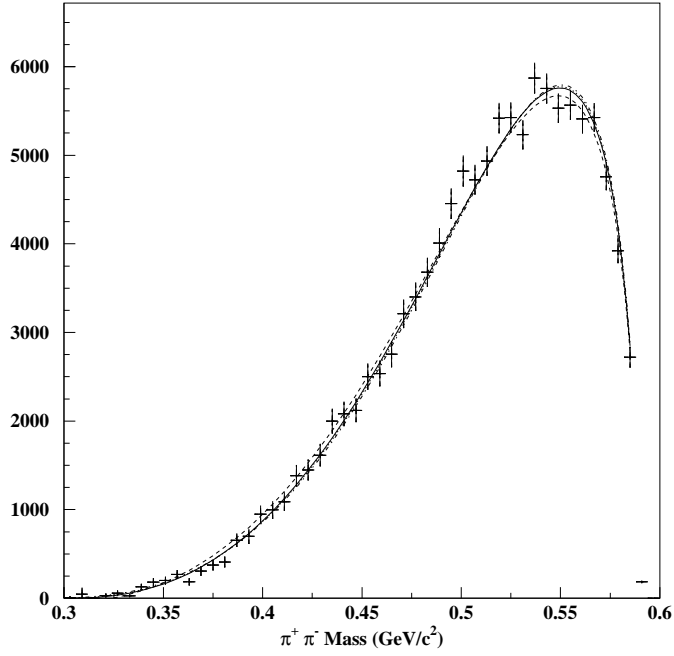}}
\caption{\label{fig:jingyun_mpipi}
Fits to the $m_{\pi \pi}$
distribution.  The points are the data corrected for
efficiency, and the curves are the fit results.  The smooth curve is
the Novikov-Shifman model (Eqn.~\ref{eqn:f1}).
The long-dashed and short-dashed curves are the T. M. Yan model with
and without higher order corrections, 
and the dash-dot
curve is the Voloshin-Zakarov model (Eqn.~\ref{eqn:voloshin}).
Three of the models are nearly indistinguishable.  The T. M. Yan model
without higher order corrections is slightly different.
The results are given in
Tables~\ref{table_kappa} and \ref{table_mpipifit}.
}

\end{figure}

\begin{figure}[!htb]
\centerline{\epsfysize 2.5 truein
\epsfbox{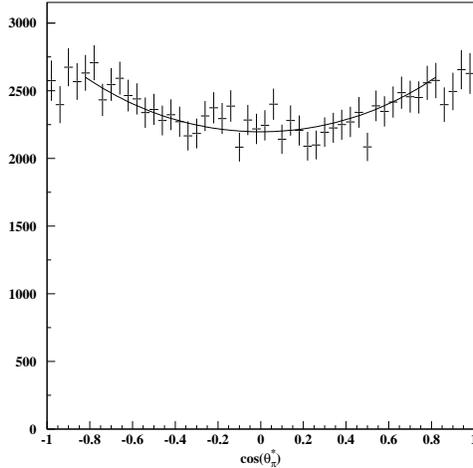}}
\caption{\label{fig:jingyun_costheta}
Fits to  $\cos \theta_{\pi}^*$ distribution. The results are given in
Tables~\ref{table_kappa} and \ref{table_mpipifit}.
The points are the data corrected for
efficiency, and the curve is the fit result using Eqn.~\ref{eqn:g1}.
}
\end{figure}

The $m_{\pi \pi}$ distribution is fit using Eqn.~\ref{eqn:f1}, as
shown in Fig.~\ref{fig:jingyun_mpipi}.
The fit yields $\kappa = 0.186 \pm 0.003$ with a $\chi^2$/DOF = 55/45.
Fitting the $\cos \theta$ distribution in the region $-0.8 < \cos
\theta_{\pi}^* < 0.8$ using Eqn.~\ref{eqn:g1} \cite{range},
we obtain the results shown in Fig.~\ref{fig:jingyun_costheta}. 
The fit yields $\kappa = 0.210 \pm 0.027$ with a $\chi^2$/DOF = 26/40.

We have also
fit the joint $\cos \theta_{\pi}^*$ and $m_{\pi \pi}$
distribution (Eqn.~\ref{eqn:dif}).  This approach does not require
integrating over one of the variables and is sensitive to any $\cos
\theta_{\pi}^*$ - $m_{\pi \pi}$ correlation.
Using this approach, we obtain a $\kappa = 0.183 \pm 0.002$ and a $\chi^2/DOF
= 1618/1482$. The results of the different fits are in good agreement
and
are summarized in Table~\ref{table_kappa}.

\begin{table}
\caption{ }
\begin{center}
\begin{tabular}{lcc}
\label{table_kappa}
Distribution           &      $\kappa$     & $\chi^2/DOF$  \\ \hline
$m_{\pi \pi}$ (Fig.~\ref{fig:jingyun_mpipi})   & $0.186 \pm 0.003 \pm 0.006$ &  55/45     \\
$\cos \theta_{\pi}^*$ (Fig.~\ref{fig:jingyun_costheta})   & $0.210 \pm 0.027 \pm 0.056$ &  26/40     \\
$m_{\pi \pi}$ vs $\cos \theta_{\pi}^*$    & $0.183 \pm 0.002 \pm 0.003$ &  1618/1482     \\
\end{tabular}
\end{center}
\end{table}

Using  Eqns.~\ref{eq:a} and \ref{eqn:dif}, where we write 
Eqn.~\ref{eq:a} in terms of S-wave and D-wave parts: $ A = A_S + A_D$, 
the ratio of the D-wave transition rate to the total rate can be
obtained
\[  R_D = \frac{\displaystyle \int dq^2 
\displaystyle \int^1_{-1} d \cos \theta (PS) | A_D |^2}
{\displaystyle \int dq^2 \int^1_{-1} d \cos \theta (PS) | A_S + A_D |^2} \]
The limits of the $q^2$ integration are $q^2_{min} = 4 m^2_{\pi}$
and $q^2_{max} = (M_{\psi(2S)} - M_{J/\psi})^2$.
For the value of $\kappa$ obtained from the joint
$\cos \theta_{\pi}^*-m_{\pi \pi}$ fit, we obtain
$ R_D = 0.184 \%$. 

\begin{figure}[!htb]
\centerline{\epsfysize 3.0 truein
\epsfbox{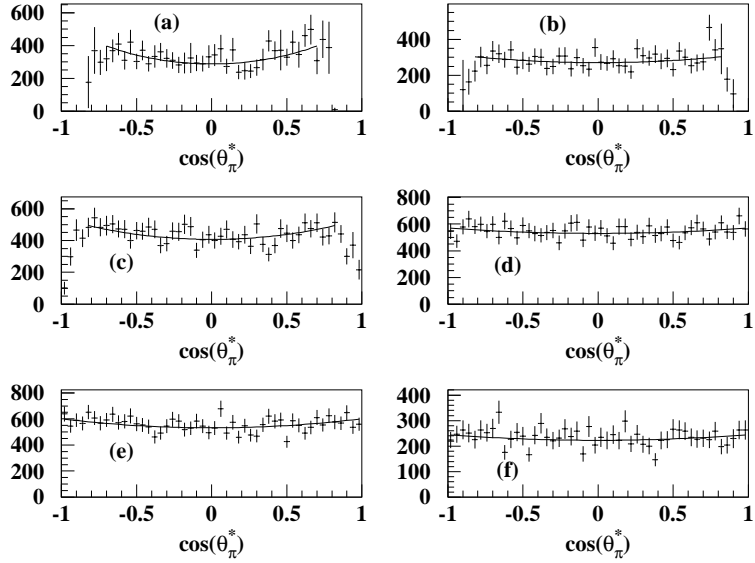}}
\caption{\label{fig:costhvsm} Fits of $\cos \theta_{\pi}^*$ using
Eqn.~\ref{eqn:pq} as a function of $m_{\pi \pi}$. The fit results
are shown in Table~\ref{tab:pq}.
{\bf (a)} $ 0.34 < m_{\pi \pi} < 0.45$ GeV/c$^2$,
{\bf (b)} $ 0.45 < m_{\pi \pi} < 0.48$ GeV/c$^2$,
{\bf (c)} $ 0.48 < m_{\pi \pi} < 0.51$ GeV/c$^2$,
{\bf (d)} $ 0.51 < m_{\pi \pi} < 0.54$ GeV/c$^2$,
{\bf (e)} $ 0.54 < m_{\pi \pi} < 0.57$ GeV/c$^2$, and
{\bf (f)} $ 0.57 < m_{\pi \pi} < 0.60$ GeV/c$^2$.
}
\end{figure}
\begin{figure}[!htb]
\centerline{\epsfysize 2.5 truein
\epsfbox{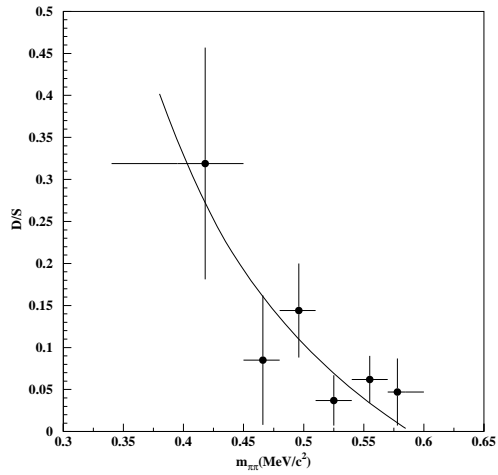}}
\caption{\label{fig:pvsm} Plot of the interference term, $\frac{D}{S}$, from
Eqn.~\ref{eqn:pq} versus $m_{\pi \pi}$.  The smooth curve is the prediction
of the Novikov-Shifman model for $\kappa = 0.183$.
}
\end{figure}

The amount of D wave as a function of $m_{\pi \pi}$ has been fit
using 
\begin{equation}
N(\cos \theta) \propto 1.0 + 2\left(\frac{D}{S}\right)\left(\cos^2 \theta - \frac{1}{3}\right) +
\left(\frac{D}{S}\right)^2 \left(\cos^2 \theta - \frac{1}{3}\right)^2 \label{eqn:pq}
\end{equation}
The last term corresponds to the amount of D-wave, while the middle term
corresponds to the interference term\cite{model}.  The results are shown in
Fig.~\ref{fig:costhvsm} and in Table~\ref{tab:pq}.
The behavior of
$\frac{D}{S}$ as a function of $m_{\pi \pi}$ is shown in
Fig.~\ref{fig:pvsm}, along with the prediction of
the Novikov-Shifman model.

\begin{table}[!h]
\caption{ Fit results to $\cos \theta_{\pi}^*$ distribution using
a $\chi^2$ fit to Eqn.~\ref{eqn:pq}. 
The fit also requires a normalization term which is not shown.
}
\begin{center}
\begin{tabular}{ccccr}
$m_{\pi \pi}$ Range~(GeV/c$^2)$~&~~$\frac{D}{S}$~~& $\chi^2$/DOF& Events \\ \hline
\label{tab:pq}
 0.34 - 0.45   & $0.319 \pm 0.097 \pm 0.098$     & 24/37 & 2016\\
 0.45 - 0.48   & $0.085 \pm 0.068 \pm 0.036$     & 29/40 & 1995 \\
 0.48 - 0.51   & $0.144 \pm 0.045 \pm 0.033$     & 33/40 & 3729 \\
 0.51 - 0.54   & $0.037 \pm 0.025 \pm 0.017$     & 35/48 & 5620\\
 0.54 - 0.57   & $0.062 \pm 0.022 \pm 0.017$     & 44/48 & 6403\\
 0.57 - 0.60   & $0.047 \pm 0.036 \pm 0.018$     & 48/48 & 2959\\
\end{tabular}
\end{center}
\end{table}

\clearpage
\subsection{The T. M. Yan and Voloshin-Zakorov Models}

Other models which describe the $m_{\pi \pi}$ invariant mass spectrum
are the T. M. Yan Model \cite{yan1} and the Voloshin - Zakarov Model
\cite{Voloshin}.
These models are also based on the color-field multiple expansion. Yan
suggests that the decay can be written as
\begin{eqnarray}
\frac{d \sigma}{d m_{\pi \pi}} & \propto & (PS) \times \left[(m_{\pi \pi}^2 -2 m_{\pi}^2)^2 
 + \frac{B}{3A}(m_{\pi \pi}^2 - 2 m_{\pi}^2)\left(m_{\pi \pi}^2 -4 m_{\pi}^2 \right.\right.\nonumber \\
& & + 2 K^2 \left.\left.\left(1 + \frac{2 m_{\pi}^2}{m_{\pi \pi}^2}\right)\right) + O\left(\frac{B^2}{A^2}\right)\right], 
\label{eqn:yan}
\end{eqnarray}
where

\[K = \frac{M_{\psi(2S)}^2 - M_{J/\psi}^2 + m_{\pi \pi}^2}{2 M_{\psi(2S)}} \]
The ratio $B/A$ is taken to be a free parameter.  The term
$O(\frac{B^2}{A^2})$ refers to higher order (HO) terms.

The Voloshin - Zakarov Model calculates the matrix element in the
chiral limit, $m_{\pi} = 0$, and then adds a phenomenological term
$\lambda m^2_{\pi}$
\begin{eqnarray}
\frac{d \sigma}{d m_{\pi \pi}} \propto (PS) \times [ m_{\pi \pi}^2 -
\lambda  m_{\pi}^2]^2. \label{eqn:voloshin}
\end{eqnarray}

\noindent The $m_{\pi \pi}$ invariant mass spectrum has been fit with these
models, as shown in Fig.~\ref{fig:jingyun_mpipi}. As can be seen, the
Novikov-Shifman and the Voloshin-Zakorov models give nearly identical fits.
The T. M. Yan model, neglecting higher order terms does not agree as well with
the data.  Including the higher order terms \cite{yan1}, however, gives a fit
result which is nearly identical to the other two models, as seen in
Fig.~\ref{fig:jingyun_mpipi}.  All the results are
summarized in Table~\ref{table_mpipifit}, along with the $\psi(2S)$
results from Argus \cite{argus}, which used $\psi(2S)$ data from Mark II. 
Argus did not fit the T. M. Yan model with the HO corrections, but the
the agreement is good for the fits they did.
 
\begin{table}[!h]
\caption{Fit Results for the $m_{\pi \pi}$ distribution.
} 
\
\begin{center}
\begin{tabular}{lcc}
Model~~&~~~BES~~~&~~Argus - MKII \cite{argus} \\  \hline \label{table_mpipifit}
Novikov-  & $\kappa = 0.186 \pm 0.003 \pm  0.006$  & $0.194 \pm 0.010$   \\
Shifman \cite{vzns}         & $\chi^2/DOF$ = 55/45          & 38/24           \\ \hline
T. M. Yan \cite{yan1} & $B/A = -0.225 \pm 0.004 \pm 0.028$    &  $-0.21 \pm 0.01$    \\
         &  $\chi^2/DOF = 84/45$         &                    \\ \hline
T. M. Yan \cite{yan1} &  $B/A = -0.336 \pm 0.009 \pm 0.019$  &       \\
(HO)     &   $\chi^2/DOF = 60/45$       &      \\ \hline
Voloshin- & $\lambda = 4.35 \pm 0.06 \pm 0.17$   &         \\
Zakorov \cite{Voloshin}  &  $\chi^2/DOF = 69/45$         &        \\
\end{tabular}
\end{center}
\end{table}

\subsection{The T. Mannel - M. L. Yan Model}
Mannel has constructed an effective Lagrangian using chiral symmetry
arguments to describe the decay of heavy excited S-wave spin-1
quarkonium into a lower S-wave spin-1 state \cite{mannel}.  Using
total rates, as well as the invariant mass spectrum from Mark II via
ARGUS \cite{argus}, the parameters of
this theory have been obtained.  More recently, M. L. Yan 
\etal \cite{mlyan} have pointed out that this model allows D-wave,
like the Novikov-Shifman model.  In this model, the amplitude can be
written \cite{mlyan}

\begin{eqnarray}
A & \propto &\{q^2 - c_1(q^2 + |\vec{q}|^2)\left(1 + \frac{2 m_{\pi}^2}{q^2}\right)
+ c_2m_{\pi}^2\} \nonumber\\
& & +\frac{3}{2}\left[c_1  |\vec{q}|^2\left(1 -\frac{4
m^2_{\pi}}{q^2}\right)\right]\left(\cos^2 \theta_{\pi}^* - \frac{1}{3}\right), \label{eqn:mannel}
\end{eqnarray}
where
\begin{eqnarray}
c_1 & = & -\frac{g_1}{3g}\left(1 +\frac{g_1}{6g}\right)^{-1} \nonumber \\
c_2 & = & 2\left(\frac{g_3}{g} - \frac{g_1}{3g} - 1\right)\left(1
+\frac{g_1}{6g}\right)^{-1} 
\end{eqnarray}
and
\[ |\vec{q}| = \frac{1}{2 m_{\psi(2S)}}[(m_{\psi(2S)}^2 - (m_{\pi \pi}
+ m_{J/\psi})^2)(m_{\psi(2S)}^2 - (m_{\pi \pi} - m_{J/\psi})^2)]^\frac{1}{2} \]
\[ q^2 = m_{\pi \pi}^2 \]

The first term in Eqn.~\ref{eqn:mannel} is the S-wave term, and the
second is the D-wave term.  Note that another constant in the
effective Lagrangian, $g_2$, has been taken to be zero since it is
suppressed by the chiral symmetry breaking scale. 
This amplitude is similar to Eqn.~\ref{eq:a} but contains an
extra term proportional to $m^2_{\pi}$.

\begin{figure}[!htb]
\centerline{\epsfysize 2.5 truein
\epsfbox{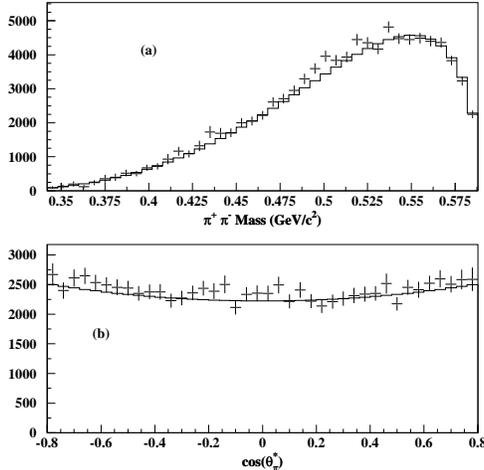}}
\caption{\label{fig:mannel2d}
Fit of the 2D $\cos \theta_{\pi}^*$ versus $m_{\pi \pi}$ distribution to
Eqn~\ref{eqn:mannel}. {\bf (a)} The 2D distribution projected in $m_{\pi \pi}$.
{\bf (b)} The 2D distribution projected in $\cos
\theta_{\pi}^*$. The points are the data corrected for
efficiency, and the histogram is the projected fit result. 
}
\end{figure}

We have
fit the joint $\cos \theta_{\pi}^*$ - $m_{\pi \pi}$
distribution using the amplitude of Eqn.~\ref{eqn:mannel} \cite{histat}, as
shown in Fig.~\ref{fig:mannel2d}.
We obtain:
\begin{eqnarray}
\frac{g_1}{g} & = & -0.49 \pm 0.06 \pm 0.13 \nonumber\\
\frac{g_3}{g} & = & 0.54 \pm 0.23 \pm 0.42\nonumber
\end{eqnarray}
with a $\chi^2/DOF = 1632/1481.$

In the chiral limit, $g_3 = 0$.  If we fit with this value for $g_3$,
we obtain
\[\frac{g_1}{g}  =  -0.347 \pm 0.006 \pm 0.007  \]
with a $\chi^2/DOF = 1632/1482$.
The results for both
cases are given in Table~\ref{table:mannel}, along with the results from
Ref.\cite{mannel} which are based on ARGUS-Mark II \cite{argus}.  The
results agree well for the  $g_3 = 0$ case.  The agreement is not as good
for the   $g_3 \neq 0$ case, but Ref.\cite{mannel} used only the 
$m_{\pi \pi}$ distribution in their fit.  In both cases, the
$\chi^2/$DOF is large, and there is no reason to prefer one fit over the other.

\begin{table}[!h]
\caption{Fit results using Eqn~\ref{eqn:mannel}. In the second fit,
$g_3$ is set to zero.
} 
\
\begin{center}
\label{table:mannel}
\begin{tabular}{lccc}
~~~~&~~~$g_1/g$~~~&~~$g_3/g$~~ &~~~$\chi^2/DOF$ \\ \hline
This Exp. &   $-0.49 \pm 0.06 \pm 0.13$ &   $0.54 \pm 0.23 \pm 0.42$
& 1632/1481 \\
Ref.~\cite{mannel}      & $-1.55 \pm 0.51$    &  $4.07 \pm 1.56$ &
0.87    \\ \hline
This Exp. &   $ -0.347 \pm 0.006 \pm 0.007$   &    0   &  1632/1482       \\

Ref.~\cite{mannel}      & $-0.35 \pm 0.03$    &  0  & 1.05   \\
\end{tabular}
\end{center}
\end{table}

\section{Systematic Errors}

The systematic errors quoted throughout this paper were
determined from the changes in the calculated results due to
variations in cuts, binning changes in the fitting
procedures, and changes due to making an additional cut to eliminate
background.  Cut variations include changing the $\cos
\theta_{\pi}$ selection from 0.75 to 0.8, changing the $pxy_{\pi}$ cut from
0.1 to 0.08 MeV/c, 
changing the $\cos \theta_{\mu}$  cut from 0.6 to 0.65, changing the
$\cos \theta_e$ cut from 0.75 to 0.7, and changing
the $m_{recoil}$ cut to $3.05 < m_{recoil} < 3.14$ .

Fitted results are sensitive to the region of the histogram used in
the fitting procedure.  The changes obtained with reasonable variations in
the number of bins used were included in the systematic error.

In addition, the events
were fitted kinematically, and a $\chi^2$ cut was made on the fitted
events. Changing the $m_{recoil}$ cut and cutting on the kinematic fit
$\chi^2$ determines the contribution to the systematic errors
due to backgrounds remaining in the event sample.

\section{Summary}

In this paper, we have studied the process $\ppj$.  We find reasonable
agreement with a simple Monte Carlo model except for the distribution
of $\cos \theta_{\pi}^*$, which is the cosine of the angle of the pion
with respect to the $J/\psi$ direction in the rest frame of the $\pi
\pi$ system.  Some D-wave is required
in addition to S-wave.

The angular distributions are compared with the general decay
amplitude analysis of Cahn. 
We find that $|M_{201}|/|M_{001}|$, which measures the amount of D-wave of
the dipion system relative to the amount of S-wave, varies between
0.12 and 0.18 and is at least two sigma from zero.  On the other hand
$|M_{021}|/|M_{001}|$, which measures the amount of D-wave of the
$J/\psi$ - $X$ system relative to the S-wave,  varies between -0.04
and 0.06 and is, in all cases, consistent with zero.
We are unable to
fit for the $\pi \pi$ phase shift angle, $\delta^0_0$.

The $m_{\pi \pi}$ distribution has been fit with the Novikov-Shifman,
T. M. Yan (with and without higher order terms), and Voloshin-Zakorov
models.  The models
give very similar fits except for the T. M. Yan model without higher order
terms, which gives a poorer fit to the data.  All fits yield a $\chi^2/DOF$
larger than one.

In addition, the Novikov-Shifman model, which is written in terms of
the parameter $\kappa$, predicts that D-wave should be present if
$\kappa$ is non zero.  Determinations of $\kappa$ based on the $\cos
\theta_{\pi}^*$ distribution and the joint $m_{\pi \pi}$ - $\cos
\theta_{\pi}^*$ distribution agree with the value obtained from the
$m_{\pi \pi}$ distribution.  The results agree well with the
measurement of Argus using Mark II data. However the fit
to the joint $m_{\pi \pi}$ - $\cos \theta_{\pi}^*$ distribution also
yields a $\chi^2/DOF$ which is larger than one.

The $\cos \theta_{\pi}^*$ distribution has been fit
to determine the amount of D-wave divided by the amount of S
wave, $\frac{D}{S}$,
as a function of $m_{\pi \pi}$. 
It is found to decrease with
increasing $m_{\pi \pi}$ in agreement with the prediction of the
Novikov-Shifman model.

Finally, we have fit our $m_{\pi \pi}$ - $\cos
\theta_{\pi}^*$ distribution using the Mannel-Yan model, which also
allows D-wave.  We find good agreement with their result
obtained in the chiral limit where $g_3 = 0$ using
the Mark II data. 


We would like to thank the staff of BEPC accelerator and the IHEP Computing
Center for their efforts.  We also wish to acknowledge useful
discussions with R. Cahn, M. Shifman, W.S. Hou, S. Pakvasa, Y. Wei,
M. L. Yan, and T. L. Zhuang.



\begin{references}
\bibitem{binbybin}
How the bin-by-bin efficiency is used depends on the type of fit used.
When the bin statistics is high, as in the 1D
histograms, the data histogram is divided by the efficiency histogram,
and the efficiency corrected  histogram is compared to theory.  When the bin
statistics is small, as in the 2D histograms, and Poisson
statistics becomes important, the theory distribution is multiplied by
the efficiency histogram, and the resulting histogram is compared with the
detected events.
\bibitem{mass_res}
The $m_{\pi \pi}$ mass resolution in this region, determined using
Monte Carlo events, is 6.4 MeV/c$^2$.  It should be noted that the
bin-by-bin efficiency correction method compensates for the effects
of resolution smearing.
\bibitem{fit_range} The range of $\cos \theta^*_X$ and $\cos
  \theta^*_{\pi}$ in this figure are reduced compared to
  Fig.~\ref{fig:mccompare} because the efficiency correction is less
  certain near the limits of the plots.  However the full range shown
  in  Fig.~\ref{fig:mccompare} is fit and used in the determination of the
  systematic errors of the fit quantities.
\bibitem{delta00} We assume that $\delta^0_0$ changes linearly as
a function of $m_{\pi \pi}$.
\bibitem{range} The fit is limited to this range because the efficiency
is less certain near the limits of the plot.  However, the full range is
used in the determination of the systematic error.
\bibitem{model} 
$S$ and $D$ have been treated as real. Treating them as
complex numbers would add a cosine of a phase angle multiplying $D/S$
in the interference term.
\bibitem{histat}For this fit, we use a larger ($\times 10$) Monte Carlo sample.
\end{references}
\end{document}